\documentclass[12pt]{tMPH2e}

\usepackage{graphicx}
\usepackage[utf8]{inputenc}

\begin{document} 
 
\title{ Critical region of D-dimensional spins: Extension and analysis of the hierarchical reference theory}

\author{Enrique Lomba$^{a}$ and Johan S. Høye$^{b}$\\
\vspace{6pt}
$^a$\em{Instituto de Química Física Rocasolano, CSIC,  Serrano 119,
  E-28006, Madrid, Spain}\\
$^{b}$\em{Institutt for Fysikk, NTNU, N-7491 Trondheim, Norway}
}

\maketitle

\begin{abstract}
The hierarchical reference theory (HRT) is generalized to spins of dimensionality $D$. Then its properties are investigated by both analytical and numerical evaluations for supercritical temperatures. The HRT is closely related to the self-consistent Ornstein-Zernike approximation (SCOZA) that was developed earlier for arbitrary $D$. Like the $D=1$ case we studied earlier, our investigation is facilitated by a situation where both HRT and SCOZA give identical results with a mean spherical model (MSM) behavior (i.e.~$D=\infty$). However, for the more general situation we find that an additional intermediate term appears. With an interplay between leading and subleading contributions, simple rational numbers, independent of $D$ ($<\infty$), are found for the critical indices.
\end{abstract}

%111111111111111111111111111111111111111111111111111111111111111111111111111111111111111111111111111111
\section{Introduction}
\label{sec1}

In a previous work we analyzed and investigated numerically the critical region of the hierarchical reference theory (HRT) for fluids, lattice gases and Ising spins \cite{hoye11}. The HRT  was introduced by Parola and Reatto as a new and accurate method to evaluate the equation of state of fluids \cite {parola85}. The HRT was inspired by the renormalization  group that was developed by Wilson and Kogut \cite{wilson74}. By numerical work it was found that the HRT gave very accurate results for the equation of state for continuum fluids. Especially, the critical region could be described well, and very good results for critical properties were obtained \cite{parola85,parola93,parola95,Parola2012}. 

Another accurate method to obtain the equation of state is the self-consistent Ornstein-Zernike approximation (SCOZA) developed by H{\o}ye and Stell \cite{hoye77,hoye84,hoye85}. With SCOZA the correlation function is assumed to be of the Ornstein-Zernike form \cite{ornstein14} that contains a free parameter to be determined by thermodynamic self-consistency such that the equation of state becomes the same both via the energy and compressibility routes. The HRT also uses the Ornstein-Zernike form with a free parameter with a corresponding self-consistency between a free energy expression and compressibility. The main difference between the HRT and the SCOZA is that the former adds smaller and smaller wave vectors to the perturbing interaction while the latter adds strength to it by lowering the temperature. It can be noted that the Ornstein-Zernike form also is the leading perturbing contribution to the correlation function by $\gamma$-ordering where $\gamma$ is the inverse range of interaction \cite{hemmer64,lebowitz65}.

Due to common features of HRT and SCOZA a unification of these theories was initiated by Reiner and H{\o}ye \cite{reiner05}. They considered the mean spherical model (MSM) and a generalized version of it, the generalized MSM (GMSM) where the exact solution was found by both the HRT and SCOZA by use of one free parameter \cite{hoye07}. These models are precisely the $D=\infty$ case of the $D$-dimensional spins to be considered in this work. However, for the usual $D=1$ case the correlation function of Ornstein-Zernike form will have two free parameters to be determined self-consistently. By the unification it was found that the resulting problem was essentially a sum of the HRT and SCOZA problems.

Analysis of the unified problem was initiated by one of the authors \cite{hoye09}. Earlier it was found from numerical work that SCOZA alone resulted in a generalized scaling with different critical indices for super- and subcritical temperatures \cite{borge98,pini98}. By analytic evaluations it was confirmed that the SCOZA critical indices were simple rational numbers \cite{hoye00}. The HRT, however, lead to usual scaling, but  from numerical evaluations  critical indices seemed to vary somewhat (except for the index of the critical isotherm, $\delta=5$, that follows from the given interaction). This variation seemed to depend upon whether a sharp or smooth cut-off was used when adding wave vectors to the interaction \cite{ionescu07,parola08,parola09}. With the unified problem there would be reason to expect that the different behaviors of these two theories might be straightened out.

By the analysis performed in Ref.~\cite{hoye09} it was found that the HRT part of the unified problem would dominate close to the critical point. But despite this it turned out that HRT somehow had to reconcile with SCOZA properties whose generalized scaling was tied to a connection between leading and subleading contributions to the critical isotherm. This situation is also present for the MSM and the GMSM. These contributions represent the connection to the mean field boundary conditions away from the critical point. It was  justified by the analysis, that the critical isotherm of the HRT also should have the same subleading contribution. Moreover, in order to obtain full scaling, instead of generalized scaling, it was concluded that HRT should  also produce an additional intermediate contribution. The leading and the two subleading  levels of  contributions were further connected via thermodynamic self-consistency. This leads to simple rational numbers for the critical indices.

In recent work the authors performed further analysis and numerical work, restricted to supercritical temperatures, so as to reveal the critical properties of the HRT   for $D=1$ \cite{hoye11}.   Within numerical accuracy, the properties of the HRT mentioned above were confirmed with the conclusion that the critical indices in standard notation for 3 dimensions are $\alpha=0$, $\delta=5$, $\beta=1/3$, $\gamma=4/3$, $\eta=0$, and $\nu=2/3$.   This seems to disagree with previous evaluations of HRT that numerically gave somewhat different numbers \cite{parola85,parola93,parola95,Parola2012}. However, in view of our analysis we see this differently. A crucial reason for different numbers for the critical indices is the assumed form of the the critical behavior. The usual assumption of a scaling behavior expressed by one power law gives one result while our analysis shows a structure of a leading level and two levels of subleading scaling functions of importance. Our numerical results that fit accurately into these functions, confirm this structure of the HRT in the critical region.  

  In Ref.~\cite{Parola2012} it was noted that the HRT in the critical region becomes equivalent to a renormalization group generator that has been studied earlier \cite{nicoll76}. Thus our results may imply that the renormalization procedure more generally also can be expected to contain the HRT structure of leading and subleading contributions.

As argued in Ref.~\cite{hoye09}, it is not ruled out that these are the exact indices for fluids and Ising spins in 3 dimensions apart from corrections of logarithmic type (see Secs.~X and XI of Ref.~\cite{hoye09}   where arguments and explanations are given in some detail). This, however, is at variance with known epsilon-expansion results \cite{fisher67, pelissetto02} where estimates are $\alpha=0.11$, $\delta=4.789$, $\beta=0.327$, $\gamma=1.237$, and $\eta=0.036$ \cite{parola09}. A series of similar results for the critical indices are given in table 6 of Ref.~\cite{pelissetto02}. Again this disagreement may be similar to the one with HRT. The presence of leading and subleading levels of scaling functions that are connected, may also here lead to simple rational numbers for the critical indices. However, this conjecture will not be investigated further here. So in this respect our results are restricted to the HRT. From our results below we find  reason to extend the arguments of Ref.~\cite{hoye09}
to $D>1$ too.  

In the present work we want to extend the HRT to D-dimensional spins, i.e.~the spin dimensionality is $D$ while the spatial dimension is still 3.   For this case the spin-spin interaction will have the spherical symmetry of $D$ dimensions, and the order parameter will be a $D$-dimensional vector. For $D=3$ this is the classical Heisenberg model.   Earlier the SCOZA was extended to D-dimensional spins \cite{hoye97}. By combining this with the unified problem for $D=1$ studied in Refs.~\cite{hoye09} and \cite{hoye11} we find it rather obvious how the dominating contribution to HRT in the critical region can be generalized to $D>1$.  The resulting equation will contain a combination of terms for the $D=1$ and $D=\infty$ cases. Such an extension of HRT has recently also been performed by Parola and Reatto \cite{Parola2012}.

  The type of analysis and numerical work of Ref.~\cite{hoye11} can be extended to this more general situation. The results we find from this study for supercritical temperatures, will show that the critical indices do not vary with $D$. Again this is not in agreement with the results already found for HRT in Table 2 of Ref.~\cite{Parola2012} where it is found that the HRT critical index $\nu$ varies with $D$. This disagreement may again in our opinion be related to the assumed type of power law behavior in the critical region as mentioned above. This is clearly demonstrated by the difference in conclusions that can be drawn from our Figs.~\ref{fig4} and \ref{fig5} below. 
Also this is not in agreement with $1/n$ or $1/D$ expansions of critical indices \cite{fisher72}. However, for large $D$ we find that there will be a rapid crossover to the MSM ($D=\infty$) behavior away from the critical point. This can be regarded as having effective critical indices that vary with $D$. But the question about the exact critical behavior of $D$-dimensional spins
is separate from the solution of its HRT problem, and it will, as mentioned above, not be investigated further here. So in this respect our results are limited to the critical properties of the HRT for such spins.

In Sec.~\ref{sec2} we establish the HRT partial differential equation for D-dimensional spins in the critical region. This equation is a generalization that combines the $D=\infty$ case considered in Ref.~\cite{hoye07} and the $D=1$ case of Refs.~\cite{hoye09} and \cite{hoye11}. This is the equation to be analyzed and numerically solved.

In Sec~\ref{sec3} the unified HRT and SCOZA problems are considered. As only terms dominant in the critical region are kept, this results in the HRT equation of Sec.~\ref{sec2} and a SCOZA equation modified by a parameter $\nu$. By combining these two equations along with the internal energy expression an equation that reveals key   properties of $\nu$, is obtained. Further we find that these properties also must be present in the HRT.  

In Sec.~\ref{sec4} it is noted that for a   suitable, and perhaps somewhat artificial,  choice of the cut-off function $L$ for wave vectors,   the parameter $\nu$ will be a constant that can be put equal to zero. The consequence of this choice is that HRT and SCOZA both will have the same solution and thus the same critical properties   where generalized scaling is present \cite{hoye00}. For this case the explicit solution for both the leading scaling function (fixed point solution) and the subleading contribution are found. The reason to study this special case is to have an explicit analytic solution to better understand the general situation where the intermediate scaling contribution will show up to enable standard scaling.  

In Sec.~\ref{sec5} we present a brief sketch of the numerical approach and analyze the numerical results. Here the cut-off function $L$ will have a more general form such that the HRT and SCOZA solutions do no longer coincide, i.e.~$\nu$ will vary. By that the HRT will have an intermediate subleading contribution that is non-zero.

Finally, in Sec.~\ref{sec6} the critical properties are considered in view of the properties of the parameter $\nu$ found in Sec.~\ref{sec3}.    Then by comparing the powers of the leading and the two subleading contributions to thermodynamic quantities,   one finds that they are linked together. In this way definite values for the critical indices are obtained.

%22222222222222222222222222222222222222222222222222222222222222222222222222222222222
\section{The HRT equation}
\label{sec2}

In this work we want to study the critical properties of an extension of the HRT to D-dimensional spins. As other details are not sought for, the HRT equation can be established on a simplified form in a straightforward way. For $D=1$ (fluids and Ising spins) this simplified   equation, which follows from Eq.~(\ref{100}) below,   becomes Eq.~(5.7) of Ref.~\cite{hoye09} or Eq.~(3) of Ref.~\cite{hoye11} which is
\begin{equation}
y_Q+\frac{\partial}{\partial m}\left(Ly_m\right)=0.
\label{21}
\end{equation}
Here $y$ is the inverse susceptibility or compressibility, $m$ is the magnetization or order parameter, and $Q$ is the lower wave-vector cut-off of the perturbing  interaction in Fourier space. The subscripts $Q$ and $m$ denote derivatives $y_Q=\partial y/\partial Q$ and $y_m=\partial y/\partial m$. Further the $L$ with scaling form $L=L(y/Q^2)$ is the cut-off function. Here we will study the situation
\begin{equation}
L=\left(\frac{Q}{\sqrt{y+Q^2}}\right)^{\lambda'}.
\label{22}
\end{equation}
With $\lambda'=2$ one has the sharp cut-off considered originally in Refs.~\cite{parola85,parola93,parola95}. However, in later works a smooth cut-off with $\lambda'=4$ was studied \cite{ionescu07,parola08,parola09}. A detailed description of the role played by smooth and sharp cutoff in the HRT approach can be found in  Ref.~\cite{Parola2012}. Since we focus upon critical properties, the scaling form (\ref{22}) will be kept for all $Q$ ($>0$). 

Eq.(\ref{21}) follows from the change in free energy   $d\Psi$ when adding wave vectors in the interval $dk=dQ$   ($<0$) to the perturbing interaction whose Fourier transform is $\tilde\psi(k)$
\begin{equation}
  d\Psi\propto \ln(1-z\tilde\psi(Q)) Q^2 dQ.                      
\label{100}
\end{equation}
With normalization $\psi(0)=1$ the $y=1-z$. This defines the HRT in its original version \cite{parola85}. Later this relation was established in view of the $\gamma$-ordering\cite{reiner05,hoye07,hoye09}. With the thermodynamic relation $y\propto\partial^2 \Psi/\partial m^2$ Eq.(\ref{100}) results in the HRT partial differential equation. However, in   Refs.~\cite{hoye07} and \cite{hoye09} it was found more convenient to differentiate Eq.(\ref{100}) twice with respect to m by which Eq.(\ref{21}) follows when restricting   $\tilde\psi(Q)=1-\rm{const}\cdot Q^2+\cdots$  to small $Q$ relevant for critical properties (with const=1 and $\lambda'=2$).
To go from Eq.~(\ref{100}) to Eq.~(\ref{21}) coefficients that are constants, are put equal to one in order to simplify since they will not influence critical properties.

When $D\rightarrow\infty$ the spin model becomes the MSM (mean spherical model) \cite{berlin52,levis52}. Both SCOZA and HRT were extended to the MSM and a generalized version of it, the GMSM (generalized MSM), in Ref.~\cite{hoye07}. For these cases the longitudinal inverse susceptibility ($y\rightarrow y_{\|}$) is replaced by the transverse one, $y_{\perp}$ 
\begin{equation}
y_{\|}=\frac{\partial}{\partial m}(my_{\perp}), \quad y_{\perp}=\frac{\beta {\cal H}}{m}
\label{23}
\end{equation}
where here $\beta=1/(k_B T)$ and ${\cal H}$ is magnetic field. The $T$ is temperature and $k_B$ is Boltzmann's constant. Again we will consider the simplified form in the critical region by which the HRT equation of Sec.~5 of Ref.~\cite{hoye07} becomes
\begin{eqnarray}
y_{\perp Q}+\frac{a}{m}L_{\perp}y_{\perp m}=0
\label{24}
\end{eqnarray}
where $L_{\perp}$ is the $L$ given by Eq.~(\ref{22}) with $y$ replaced by $y_{\perp}$. The $a$ is merely   an adjustable parameter that does not change the properties of the GMSM, as commented below Eq.~(\ref{100}). We found it convenient to use the value $a=\sqrt{5}$ below.  

 Eq.~(\ref{24}) is a first order partial differential equation and it can be solved in a straightforward way by use of the equation for its characteristics \cite{Mattheij2005}. Its general solution for $D=\infty$, as found in Ref.~\cite{hoye07}, is  
\begin{eqnarray}
y_{\perp}=C_1, \quad aJ_{\perp}+\frac{1}{2}m^2=C_2
\label{26}
\end{eqnarray}
where   $J_{\perp}= J_{\perp}(y_{\perp},Q)$, $J_{\perp}(0,0)=0$, and
\begin{equation}
L_{\perp}=-\frac{\partial J_{\perp}}{\partial Q}.
\label{27}
\end{equation}
The mean field type boundary condition can be ($Q\rightarrow\infty$)
\begin{eqnarray}
y_{\perp}=m^2-t+\rm{const.}
\label{28}
\end{eqnarray}
($\rm{const.}=J_{\perp}(y_\perp,Q\rightarrow\infty)$) by which the solution   of Eq.~(\ref{24}) becomes the combination $C_2=C_2(C_1)=C_1/2$ of solution (\ref{26}) or  
\begin{equation}
m^2=-2aJ_{\perp}+y_{\perp}+t
\label{29}
\end{equation}
where  $t$ is the deviation from the inverse critical temperature. When $Q\rightarrow 0$ one will find $J_{\perp}\propto \sqrt{y_{\perp}}$.

The HRT for arbitrary spin dimensionality $D$ will be a combination of the two situations considered above. Then the free energy will get separate contributions from the longitudinal and transverse parts of the spins. With one contribution from the former there will be $D-1$ contributions from the latter. So with $y\rightarrow y_{\|}$ and $L\rightarrow L_{\|}$ in Eq.~(\ref{21}) the resulting equations can be written (in analogy to the SCOZA equations of Ref.~\cite{hoye97}) 
\begin{eqnarray}
\nonumber
y_{\| Q}+(1-\gamma')\frac{\partial}{\partial m}(L_{\|}y_{\| m})+\gamma' a\frac{\partial}{\partial m}(L_{\perp} y_{\perp m})=0\\
y_{\perp Q}+(1-\gamma')\frac{1}{m}(L_{\|}y_{\| m})+\gamma' a\frac{1}{ m}(L_{\perp} y_{\perp m})=0
\label{30}
\end{eqnarray}
where spin dimensionality $D$ will be
\begin{equation}
D=1+\frac{\gamma' a}{1-\gamma'}.
\label{31}
\end{equation}
In Eq.(\ref{30})   the number of transverse components versus the longitudinal one is reflected in the ratio   $D-1=\gamma' a/(1-\gamma'$), i.e. Eq.~(\ref{31}). The rescaling of coefficients is conveniently made such that Eqs.(\ref{21}) and (\ref{24}) are recovered for   $\gamma'=0$ and $\gamma'=1$ respectively.

  Eq.~(\ref{30}) is consistent with the HRT equations already established in Ref.~\cite{Parola2012}. These are equations (62) and (67) of the reference where different notation is used. This becomes more obvious when further comparing with its Eq.~(17) for $D=1$ and its Eqs.~(39) and (44) for sharp ($\lambda'=2$) and smooth ($\lambda'=4$) cut-offs respectively.  
Eq.~(\ref{30}) is the one to be analyzed and solved numerically for supercritical temperatures.

%33333333333333333333333333333333333333333333333333333333333333333333333333333333
\section{Unified HRT and SCOZA}
\label{sec3}

  To be able to reveal the HRT properties present in Eq.~(\ref{30}) with leading and subleading contributions, we find it necessary to do this in an indirect way. To do so we will consider the unified HRT and SCOZA problem and utilize SCOZA properties.  
The SCOZA problem for D-dimensional spins has been considered earlier by H{\o}ye and Stell \cite{hoye97}. This problem was also solved numerically  \cite{pini02} for $D=3$. A special feature of the numerical results for this case is that the isotherms are horizontal at phase coexistence with a mean field type curve of coexistence with critical index \cite{pini02}, $\beta=1/2$. This clearly differs from best estimates $\beta\approx 0.33$ and the SCOZA value $\beta=0.35$ for $D=1$; but horizontal isotherms at coexistence are correct.

Like it is for $D=1$, there are SCOZA equations that are consistent with the HRT equations   for $D>1$   of the preceding problem. They will follow from modification of Eq.~(28) of Ref.~\cite{hoye97} as
\begin{eqnarray}
y_{\|\beta}+1+\frac{\partial}{\partial m}[(1-\gamma')J_{\|}^\prime y_{\| m}+\gamma' aJ_{\perp}^\prime y_{\perp m}]+\nu_{mm}=0
\nonumber\\
y_{\perp \beta}+1+\frac{1}{ m}[(1-\gamma')J_{\|}^\prime y_{\| m}+\gamma' aJ_{\perp}^\prime y_{\perp m}+\nu_{m}]=0
\label{32}
\end{eqnarray}
Here the subscript $\beta$ means differentiation with respect to the inverse temperature $\beta$, and the prime on $J$ means differentiation with respect to $y$, i.e.~$J^\prime=\partial J/\partial y$. Further the   $J_\|=J_\|(y_{\|},Q)$ is the same function of $y_\|$ as $J_\perp$ is of $y_\perp$ by which also $L_\|=-\partial J_\| /\partial Q$  like Eq.(\ref{27}). The parameters   $\gamma'$  and $a$ are those of Eq.(\ref{30}) to   be consistent.    Further the parameter $\nu$ has been added. This will follow   if a unification of the HRT and SCOZA problems is performed in analogy with the $D=1$ case \cite{hoye09}.   The main change will be to consider a general $\gamma'$ in expressions (\ref{33}) below, not just $\gamma'=0$ or $\gamma'=1$ (for $D=\infty$). It may be noted that the parameter $a$ may be equal to one, but as mentioned we have chosen to use $a=\sqrt{5}$ below for convenience for the case with coinciding SCOZA and HRT solutions and hence also in the numerical work. 

Eq. (\ref{32}) expresses consistency between compressibility and internal energy as given by Eq.(\ref{34}) below.  The configurational internal energy, $X$, is given by Eq.(\ref{33}) below. The expression for $X$ follows from the assumed Ornstein-Zernike form of the   spin correlation function with Fourier transform   $\tilde\Gamma(k)$.   It can be written as $\tilde\Gamma (k)= \nu/(1-z\tilde\psi(k)) \rightarrow \nu/(y+Q^2) $ (for small $y$) with two free parameters $z$ and   $\nu$ ($\sim 1$) where $y=1-z$    (with $\tilde\psi(0)=1$) \cite{hoye07}.   In the present case with D-dimensional spins this is understood to be generalized by the extension $y\rightarrow y_\|, y_\perp$ and likewise with   $\Gamma$, $z$, and $\nu$. Then for a small change of  $\nu$ close to the critical point one can expand the resulting contribution to $J$ to linear order in this change. There will be contributions from both the transverse and longitudinal parts. These sum up to the $\nu$ in Eq.(\ref{33}), where the $\nu$ now is redefined to represent  the   change due to the contribution to $X$ from all these changes.

Again if one follows the derivations of the unification of HRT and SCOZA, one will find that the resulting equation mainly will be a sum of the SCOZA and HRT problems \cite{hoye07,hoye09,hoye11}. Near the critical point the HRT part will again dominate, and the direct influence of the parameter $\nu$ on the HRT part can be neglected, and Eqs.~(\ref{30}) and (\ref{32}) will be the result. Although the $\nu$ does not appear in Eq.~(\ref{30}) one can still use it   indirectly via SCOZA   to draw conclusions about properties of the solution of this equation since $\nu$ will approach $\nu=\rm{const.}$ ($=0$) away from the critical point   \cite{hoye09,hoye11}.  

By unification of SCOZA and HRT  there are two parameters $z$ and $\nu$ (for $D=1$) in the  Ornstein-Zernike form of the   spin correlation function,   given in the text   above,   to be determined self-consistently. General expressions for this unification were worked out in   Refs.~\cite{hoye07} and \cite{hoye09} to obtain the equations that determine $z$ and $\nu$.  There by order of magnitude analysis it was found that by approach to the critical point the unified problem simplified to   Eqs.~(\ref{33}) and (\ref{34}) below. The $\nu$-dependence of  $Y’$ can then be neglected \cite{hoye11,hoye09} by which the unified problem reduces to the HRT. Thus the resulting critical properties we obtain for the HRT are also those of the unified problem. But by our approach it is required that both the SCOZA and HRT parts are combined, as done below, to be able to conclude about the properties of both $\nu$ and HRT.

It is possible to obtain an equation for $\nu$ from which some of its properties can be deduced. Then with 
\begin{eqnarray}
\nonumber
X&=&\frac{1}{2}m^2+(1-\gamma')J_{\|}+\gamma' aJ_{\perp}+\nu\\
Y^\prime&=&\frac{\partial Y}{\partial m}=(1-\gamma')L_{\|} y_{\| m}+\gamma' a L_{\perp} y_{\perp m},
\label{33}
\end{eqnarray}
Eqs.~(\ref{30}) and (\ref{32}) can be written as
\begin{eqnarray}
\nonumber
&  y_{\| Q}&+Y^{\prime\prime}=0,\quad y_{\perp Q}+\frac{1}{m}Y^{\prime}=0\\
&  y_{\| \beta}&+X^{\prime\prime}=0,\quad y_{\perp \beta}+\frac{1}{m}X^{\prime}=0.
\label{34}
\end{eqnarray}
Here $Y$ is a function $Y=Y(y_{\|},y_{\perp},Q)$ such that $L_i$ ($i=\perp,\|$) are the partial derivatives $L_i=\partial Y/\partial y_i$. Further
the $X$ and $Y$ must be derivatives of the same free energy function $\Psi$ such that $X=\partial\Psi/\partial\beta$ and  $Y=\partial\Psi/\partial Q$; so by that
\begin{equation}
\frac{\partial X}{\partial Q}=\frac{\partial Y}{\partial\beta},
\label{35}
\end{equation}
or by use of Eq.~(\ref{33}) ($L=-\partial J/\partial Q$, $Y=(1-\gamma')\int L_{\|}\,dy_{\|} + \gamma' a\int L_\perp \,dy_\perp$) 
\begin{eqnarray}
 (1-\gamma')(-L_{\|}+J_{\|}^\prime y_{\| Q})+\gamma' a(-L_\perp +J_\perp^\prime y_{\perp Q})+\nu_Q=(1-\gamma')L_\| y_{\|\beta}+\gamma' aL_\perp y_{\perp\beta}.
\label{36}
\end{eqnarray}
This equation is the extension to $D>1$ of Eq.~(19) of Ref.~\cite{hoye11}.
Now $y_{iQ}$ and $y_{i\beta}$ ($i=\|,\perp$) are substituted by the derivatives of $X$ and $Y$. With $X$ and $Y^\prime$ given by Eq.~(\ref{33}) we have
\begin{eqnarray}
\nonumber
X^\prime&=&m+(1-\gamma')J_\|^\prime y_{\| m}+\gamma' a J_\perp^\prime y_{\perp m}+\nu_m\\
X^{\prime\prime}&=&1+(1-\gamma')(J_\|^{\prime\prime} y_{\| m}^2+J_\|^\prime y_{\| mm})+\gamma' a (J_\perp^{\prime\prime} y_{\perp m}^2+J_\perp^\prime y_{\perp mm})+\nu_{mm}
\label{37}\\
Y^{\prime\prime}&=&(1-\gamma')(L_\|^\prime y_{\| m}^2+L_\|  y_{\| mm})+\gamma' a
 (L_\perp^\prime y_{\perp m}^2+L_\perp y_{\perp mm}) 
\nonumber
\end{eqnarray}
where the primes on $J$ and $L$ mean partial derivatives with respect to $y_\|$  or $y_\perp$. Via Eq.~(\ref{34}) expressions (\ref{37}) are then inserted in Eq.~(\ref{36}) to obtain the following equation for $\nu$
\begin{equation}
\nu_Q+(1-\gamma')L_\|\nu_{mm}+\gamma' a \frac{1}{m}L_\perp \nu_m+A=0
\label{38}
\end{equation}
\begin{eqnarray}
\nonumber
A&&=(1-\gamma')^2(J_\|^{\prime\prime}L_\|-J_\|^\prime L_\|^\prime)y_{\| m}^2\\
&&+(1-\gamma')\gamma' a\left[(J_\perp^{\prime\prime}L_\|-J_\|^\prime L_\perp^\prime)y_{\perp m}^2+(J_\|^\prime L_\perp-J_\perp^\prime L_\|)\left(\frac{1}{m}y_{\| m}-y_{\perp mm}\right)\right].
\label{39}
\end{eqnarray}

For the MSM and GMSM ($\gamma'=1$) $A=0$ by which $\nu=\rm{const.}$ ($=0$), consistent with the SCOZA and HRT solutions that are the exact   ones \cite{hoye07}   as rederived in Sec.~\ref{sec2}. However, for $\gamma'<1$  (and $\lambda'>1$)   the $A\neq 0$. As argued in Refs.~\cite{hoye09} and \cite{hoye11} it then follows from Eqs.~(\ref{38}) and (\ref{39}) that the $\nu$ will contain only the same scaling terms as $y_\|$, $y_\perp$, and $X$. To see this,    consider the scaling form given by Eq.~(24) of Ref.~\cite{hoye11} (for critical index $\delta=5$, i.e.   free energy $\Psi\sim m^6$, $Q\sim m^2$)
\begin{eqnarray}
\nonumber
X&=&\frac{m^6}{t}(X_0+m^\lambda X_1+m^2 X_2+\cdots)\\
y&=&m^4(Z_0+m^\lambda Z_1+m^2 Z_2+\cdots)
\label{40}
\end{eqnarray}
where $X_i$ and $Z_i$ ($i=0,1,2$) are functions of scaled magnetization (or deviation from critical density) $z=m/Q^{1/2}$ and scaled deviation from critical temperature $\tau=t/Q^{\lambda_1}$ ($y=y_\|, y_\perp$). Like it was done in Refs.~\cite{hoye09} and \cite{hoye11} expression (\ref{33}) for $X$ is used in Eq.~(\ref{34}) for $y_{\|\beta}$ and/or $y_{\perp\beta}$.   Then, since the equation is non-linear,   terms of various orders of expression (\ref{40}) for $X$ and $y$ will be coupled together. As $J(y)\sim -\sqrt{y}$, and since the $m^2$ terms in $X$ of Eq.~(\ref{33}) should cancel, the leading term   and the subleading one in $X^{\prime\prime}$ from this equation when expanded, will be $\sim m^\lambda$   and $m^2$ respectively. In this way, when comparing with Eq.~(\ref{40})   ($X''=y_\beta\propto y/t$), we will have $m^4/t\sim m^\lambda$ and $m^{4+\lambda}/t\sim m^2$.
(Without cancelation of $m^2$ terms in Eq.~(\ref{33}), the internal energy $\propto X$ would be the mean field one.)  
This requires $\lambda=1$ and thus $t\sim m^3$; so $\lambda_1=3/2$. It is then assumed that the powers of the   coefficients $m^2$ of the $X_2$ and $Z_2$ terms are correct as they are the ones of both SCOZA and the GMSM, and as argued in Refs.~\cite{hoye09} and \cite{hoye11} they should also be the ones of the HRT. Eq.~(\ref{38}) for $\nu$ must then imply that $\nu$ will have the same scaling terms due to expression (\ref{39}) for $A$ that can be expanded in the terms of (\ref{40}) for $y$. The leading order of $A$ follows from $L\sim 1$, $J\sim\sqrt{y}$, $J^{\prime\prime}\sim J/y^2$ etc.~by which ($y\sim Q^2$) $A\sim 1$ or $\nu\sim m^2$. Thus with boundary condition $\nu=\rm{const.}$ ($=0$) for large $Q$, a non-zero solution for $\nu$ must produce the same scaling terms as contained in  $A$. In this way, inclusion of $\nu$ in the expression for $X$ will not produce new scaling terms in the solution for $y$.

  Eq.~(\ref{40}) for $y$ may express $m^2$ in terms of functions $Y_i$ ($i=0,1,2$) of the scaling variables $y/Q^2$ and $t/Q^{3/2}$ ($\lambda'=3/2$) as
\begin{equation}
m^2=y^{1/2}(Y_0+y^{1/4}Y_1+y^{1/2}Y_2+\cdots).
\label{101}
\end{equation}
This form fits directly into the mean field boundary condition (\ref{28}) which is equivalent to the initial condition for $\phi^4$ theory Eq.~(45) of Ref.~\cite{Parola2012}. Thus in Ref.~\cite{hoye09} the HRT equation was transformed into an equation for $u=m^2$ in terms of $y$ and $Q$ by which its solution in view of boundary condition (\ref{28}) could be discussed. Since expression (\ref{101}) links the critical behavior of the leading scaling function $Y_0$ directly to the boundary condition via its subleading contributions  $Y_1$ and $Y_2$, the critical properties follow as explained above.

For $Q=0$ the scaling function functions $Y_i$ will be functions of the scaling variable $w=t/y^{3/4}$. And by expansion to fit numerical data for $y=y_\perp$, we used
\begin{equation}
Y_0=a+cw+fw^2, \quad Y_1=a_1+dw,\quad Y_2=b+dg w
\label{102}
\end{equation}
where the $a$, $c$, $f$, $a_1$, $d$, and $g$ are the coefficients that are determined by the numerical evaluations.

%4444444444444444444444444444444444444444444444444444444444444444444444444444444444444444
\section{Coinciding HRT and SCOZA solutions}
\label{sec4}

If $A=0$ above, the HRT and SCOZA must coincide since then $\nu=0$. This is already the case for the MSM and GMSM \cite{hoye07} with $\gamma'=1$. However, this is also the case when ($L=L_\|,L_\perp$ etc.) \cite{hoye11}
\begin{equation}
J^\prime\propto L
\label{41}
\end{equation}
where   $J'=\partial J/\partial y$  (as in Eq.~(\ref{32})). With expression (\ref{22}) for $L$ this is so for $\lambda^\prime=1$, i.e.
\begin{equation}
L=\frac{Q}{\sqrt{y+Q^2}}
\label{42}
\end{equation}
which with relation (\ref{27}) means
\begin{equation}
J=-\sqrt{y+Q^2}.
\label{43}
\end{equation}

  The consequence of the coincidence is that the HRT properties must be those of SCOZA with its generalized scaling where super- and subcritical indices are different \cite{borge98,hoye00}. A reason to study this somewhat artificial cut-off $\lambda'=1$ is that it will give a reference to better understand the behavior for $\lambda'>1$ which is the one of interest with regular scaling present. So numerical results, except Fig.~1, are for $\lambda'>1$, i.e.~only $\lambda'=2$ is used as various values of $\lambda'$ up to $\lambda'=4$ (smooth cut-off) were investigated numerically in Ref.~1 to conclude that the HRT supercritical indices remained unchanged for $D=1$.  

When   HRT and SCOZA   coincide it turns out that it is possible to find the fixed point solution and the leading correction to it analytically. In Ref.~\cite{hoye11} the roles of $m^2$ and $y$ as independent and dependent variables were interchanged to solve the problem. This is not so obvious with both $y_\perp$ and $y_\|$ present. However, we may expect the solution to be of similar form. The fixed point solution for $t=0$ can then, with Eqs.~(\ref{33}) and (\ref{34}), be found from $X^{\prime\prime}=0$ since then the SCOZA $y_{\|\beta}=0$. Thus  $X=\rm{const}=0$ means
\begin{equation}
\frac{1}{2}m^2+(1-\gamma')J_\|+\gamma' aJ_\perp=0.
\label{43a}
\end{equation}
By closer inspection in view of   Eq.~(\ref{24})  the solution of this equation will be ($a=\sqrt{5}$)
\begin{equation}
m^2=-2J_\|=2\sqrt{y_\|+Q^2}=-2aJ_\perp=2a\sqrt{y_\perp+Q^2}.
\label{44}
\end{equation}
The leading correction to this fixed point solution will be similar to the one of Eq.~(32) in Ref.~\cite{hoye11}. With Eqs.~(\ref{44}) and (\ref{23}) we thus find
\begin{eqnarray}
\nonumber
y_\perp+Q^2&=&\frac{1}{20}m^4-cm^2(a_2 m^4+t-Q^2)+\cdots\\
y_\|+Q^2&=&\frac{1}{4}m^4-cm^2(7a_2 m^4+3t-3Q^2)+\cdots
\label{45}
\end{eqnarray}
where $c$ is an arbitrary coefficient while $a_2$ will depend upon $\gamma'$. With this one finds
\begin{eqnarray}
\nonumber
-J_\|&=&\frac{1}{2}m^2-c(7a_2m^4+3t-3Q^2)+\cdots\\
-aJ_\perp&=&\frac{1}{2}m^2-5c(a_2m^4+t-Q^2)+\cdots. 
\label{46}
\end{eqnarray}
Via Eq.~(\ref{33}) this can be inserted in  Eq.~(\ref{34}) for $y_{\|\beta}$ or $y_{\perp \beta}$. In both cases one finds
\begin{equation}
1=28(1-\gamma')a_2+20\gamma' a_2 \quad \mbox{or} \quad a_2=\frac{1}{4(7-2\gamma')}.
\label{48}
\end{equation}
 
For $\gamma'=0$ result (32) of Ref.~\cite{hoye11} is obtained while for $\gamma'=1$ the MSM result (\ref{29}) is recovered. Likewise it can be shown that the HRT equations for $y_{\| Q}$ and $y_{\perp Q}$ in Eq.~(\ref{34}) are satisfied too by noting that with relation (\ref{41}) one can replace $L$ with
\begin{equation}
Ly_m=-2QJ^\prime y_m=-2Q\frac{\partial J}{\partial m}.
\label{49}
\end{equation}

  With solution (\ref{45}) one has the SCOZA supercritical indices $\gamma=2$, $\alpha=-1$, and $\delta=5$, which are the ones of the MSM and GMSM. (The $\alpha'=-1$ corresponds to specific heat \quad $\rm{const.}-\rm{const.}|t|$.) However, evaluation of the phase equilibrium for $D=1$ gave the SCOZA subcritical indices $\gamma'=1.4$, $\alpha'=-0.1$, and $\beta=0.35$, i.e.~generalized scaling \cite{borge98,hoye00}.  

  The reason for the SCOZA behavior is that the leading $m^4$ term in Eq.~(\ref{45}) connects to the subleading one where $t$ is formally present ($Q=0$). Then one might try to modify SCOZA somehow to possibly obtain  regular scaling by which one will notice that this would require the presence of an intermediate term \cite{hoye09}. In the unified problem the parameter $\nu$ of Eqs.~(\ref{32}) and (\ref{33}) can produce such a term. This again leads to the scaling form (\ref{40}) which again leads to regular scaling.  

  Since HRT does not give an exact solution to the statistical mechanical problem the results will vary with the type of cut-off used. But according to our analysis and numerical results the critical indices will stay unchanged (for $\lambda'>1$). Instead the coefficients of Eq.~(\ref{102}) will vary such that for $\lambda'=1$ one has $c=f=a_1=g=0$, and one is left with the SCOZA or MSM type solutions (\ref{45}) or (\ref{29}) in the critical region. With this the remaining $t$-dependence scales with $y^{1/2}$ (or $m^2$) with scaling variable $t/y^{1/2}$ in the MSM. In SCOZA, however, the $t$ also scales with $m^4\propto y$ with scaling variable $t/m^4$ which lead to generalized scaling and subcritical index $\beta=0.35$ \cite{hoye00}. The reason for this is that the non-linear SCOZA equation connects the leading and subleading contributions. To have regular scaling it was realized that an intermediate contribution is required \cite{hoye09}. As argued in Sec.~\ref{sec3} this contribution is produced due to the HRT part of the unified problem by which the two approaches are reconciled. In Sec.~\ref{sec6} the supercritical properties are investigated in more detail.  

%5555555555555555555555555555555555555555555555555555555555555555555555555555555555555555555
\section{Numerical study}
\label{sec5}
\subsection{Numerical method}
For convenience we will solve the first of Eqs. (\ref{30}), which can be rewritten as 
\begin{equation}
y_{\|Q} + (1-\gamma')\left(y_{\|m}L_{\|m}+L_\|y_{\|m,m}\right)+\gamma' \sqrt{5}\frac{\partial}{\partial m}\left(L_\perp y_{\perp m}\right) = 0
\label{yq}
\end{equation}
In order to solve this equation in terms of $y_\perp$ one simply has to integrate Eq. (\ref{23}), so that $y_\perp$ can be expressed in terms of $y_\|$. Hence
\begin{equation}
y_\perp(m) = -\frac{1}{m}\int_m^0 y_\|(m') dm' \;\;\; \mbox{with}\;\; m < 0,
\label{yperp}
\end{equation}
where we have explicitly expressed the $m$-dependence of $y_\perp$, focusing on the $m<0$ region, since $y_\perp(m)=y_\perp(-m)$ and the same applies to $y_\|$. Additionally, 
\begin{equation}
y_\perp(0) = y_\|(0)\;\;\mbox{and}\;\;y_{\perp m}(0) = y_{\|m}(0)=0, 
\label{yp0}
\end{equation}
Now, Eq.~(\ref{yq}) can be solved using and implicit Euler method as in Ref.~\cite{hoye11}, by which once the equation is discretized we get
\begin{eqnarray}
y_\|(Q-\Delta Q,m) & = & y_\|(Q) +\frac{ (1-\gamma')\Delta Q}{(\Delta m)^2}\left[\frac{L_{y\|}(Q,m)}{4} (y_\|(Q,m+\Delta m)-y_\|(Q,m-\Delta m))^2\right.\nonumber\\
&+& \left. L_\|(Q,m)(y_\|(Q,m+\Delta m)-2y_\|(Q,m)+y_\|(Q,m-\Delta m))\right]\nonumber\\
&+& \frac{\sqrt{5}\gamma'\Delta Q}{2\Delta m} \left[L_\perp (Q,m+\Delta m)y_{\perp m}(Q,m+\Delta m)\right.\nonumber\\
&-&\left. L_\perp (Q,m-\Delta m)y_{\perp m}(Q,m-\Delta m)\right]
\label{numsol}
\end{eqnarray}
with $y_{\perp m}(Q,m) = (y_{\perp}(Q,m+\Delta m)-y_\perp(Q,m-\Delta m))/(2\Delta m)$ and
\begin{equation}
y_\perp(m) = - \frac{1}{m} \sum_{i=i_m}^0 \frac{1}{1+\delta_{0,i}+\delta_{i_m,i}}y_\|(i\Delta m) \Delta m\;\;\mbox{with}\;\; m = i_m\Delta m \le 0.
\label{ypm}
\end{equation}
This last expression is nothing but the trapezoidal rule applied to (\ref{yperp}). As mentioned before $L_\| = L(y_\|)$ and $L_\perp = L(y_\perp)$ with $L$ given by Eq.(\ref{22}), and 
\[ L_{y\|} = \left.\frac{\partial L}{\partial y}\right|_{y=y_\|} \]
Now,   one starts from the boundary condition at large $Q$ (i.e. the mean field limit) 
\begin{equation}
y_\| (m) = a' m^2 + b'.
\label{mf}
\end{equation}
Equation (\ref{numsol}) can be iterated backwards   (i.e.~$\Delta Q<0$) using (\ref{yp0}), (\ref{ypm}), and the symmetry of $y$ with respect to $m=0$ all the way down to $Q=0$. As in Ref.~\cite{hoye11} the stability condition\cite{Mattheij2005} 
\[  \frac{|\Delta Q|}{(\Delta m) ^2} \le \frac{1}{2} \]
must be preserved. In our calculations we have used $\Delta m = 0.005$ and $|\Delta Q| = 3\times 10^{-6}$. In all cases we have set $a' = 0.1$ in Eq.~(\ref{mf}). The quantity   $b'$ (+const.)   is proportional to the temperature. In order to keep the solution stable, we have also kept fixed both $y_\|(m_{min})$ and $y_{\|m}(m_{min})$ at the boundary, where in this case $m_{min} = -2$. 

\subsection{Results}
\begin{figure}[H]
\includegraphics[width=17cm,clip]{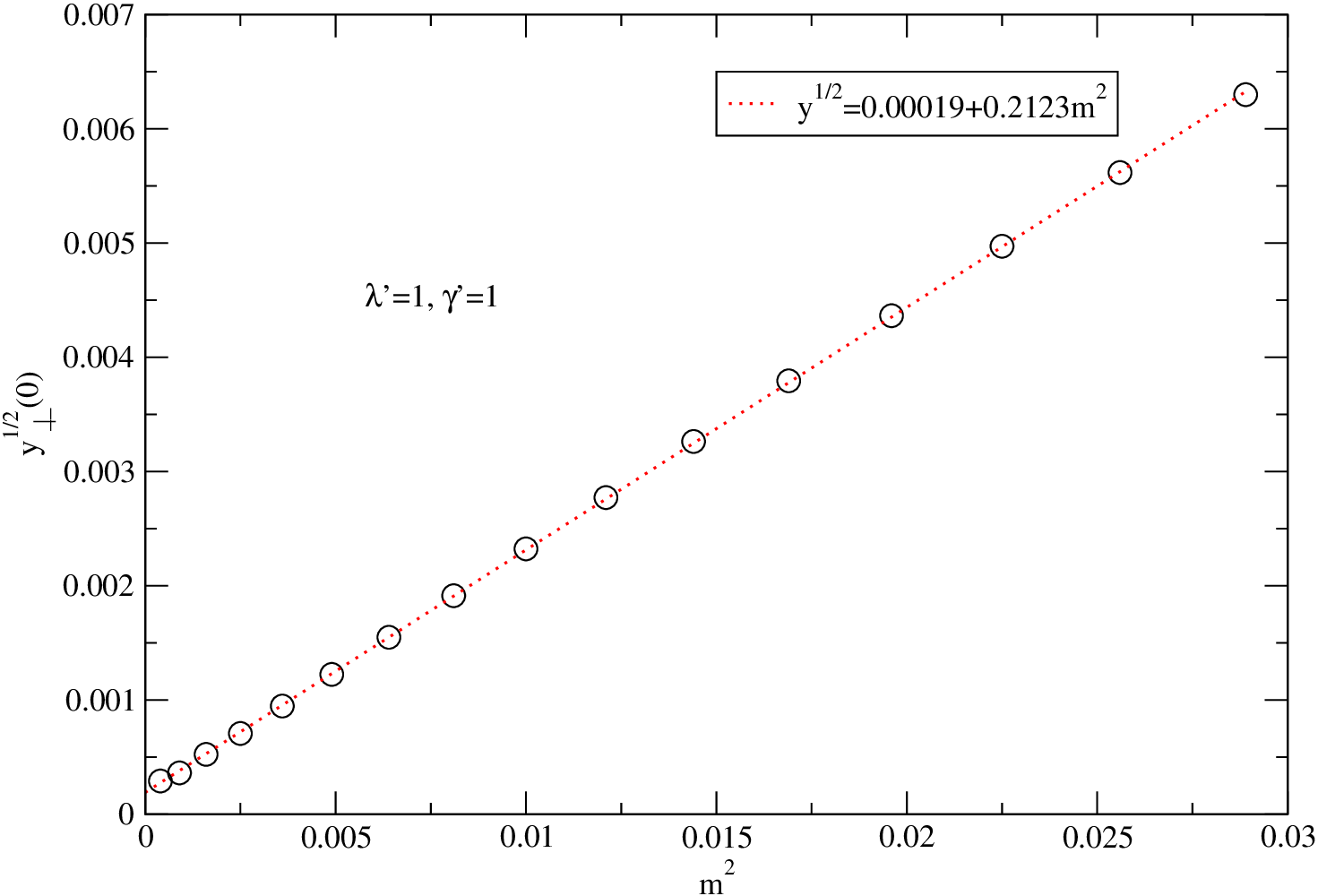}
\caption{Scaling of $y_{\perp}(0)$ vs $m$ for the spherical model case ($\gamma'=1$) when $\lambda'=1$ in Eq.(\ref{22}). Symbols correspond to the numerical solution and the line to the linear regression fit.   (The $y$ on this and the other figures below is simplified notation for $y_\perp$).}
\label{fig1}
\end{figure}

As a test of the solution procedure we have first considered the spherical model limit  ($\gamma' = 1$) for the case $\lambda'=1$. In this case we know from Eqs. (\ref{26}) and (\ref{27}) that the exact dependence
\begin{equation} 
 y_{\perp}^{1/2} (0) = \frac{1}{2\sqrt{5}} m^2 + \ldots
\label{yl1g1}
\end{equation}
must be satisfied. In Fig.~\ref{fig1} we plot the results of the numerical solution of (\ref{numsol}), where $y_{\perp}^{1/2} (0)$ is seen to scale linearly with $m^2$ with a slope close to that of the exact expression (\ref{yl1g1}). The difference can be attributed to the deviation from the true critical temperature, since due to numerical roundoff errors the lowest value of $y_\perp$ that can be reached is $\approx 10^{-7}$. 

  In Ref.~\cite{hoye11} various values of $\lambda'$ were investigated numerically for $D=1$ with the conclusion that critical properties did not change (except for $\lambda'=1$). Thus
following the analysis of Ref.~\cite{hoye11}, we focus here on the HRT sharp cut-off results ($\lambda'=2$), which qualitatively behave similarly to those of the smooth cutoff ($\lambda'=4$). By extrapolation to $y_\perp(0) = 0$, we get an estimate of $b_c'$ and $m_c$. The $b_c'$ is the value of the temperature parameter at the critical point, and $m_c$ is the magnetization per spin along the critical isotherm. As in Ref.~\cite{hoye11} and discussed in connection with Eq.~(\ref{40}), we seek for an expansion of a leading (first two terms) and two subleading contributions as given by Eqs.~(\ref{101}) and (\ref{102}) 
\begin{equation}
m_c^2-m^2 = f t^2/y_\perp +ct/y_\perp^{1/4} +dt(1+gy_\perp^{1/4})+ \ldots 
\label{fit1}
\end{equation}
with $t=b_c'-b'$. The $m_c^2$ is the critical isotherm and is the same as expression (\ref{eosDs}) for $m^2$ below taken at $t=0$, i.e.~critical isotherm. By evaluation of the difference (\ref{fit1}) numerically the $t$-dependence is found more accurately, i.e. all terms of Eq.~(\ref{102}) can be included, compared to direct use of expression (\ref{eosDs}) alone.  

\begin{figure}[H]
\includegraphics[width=17cm,clip]{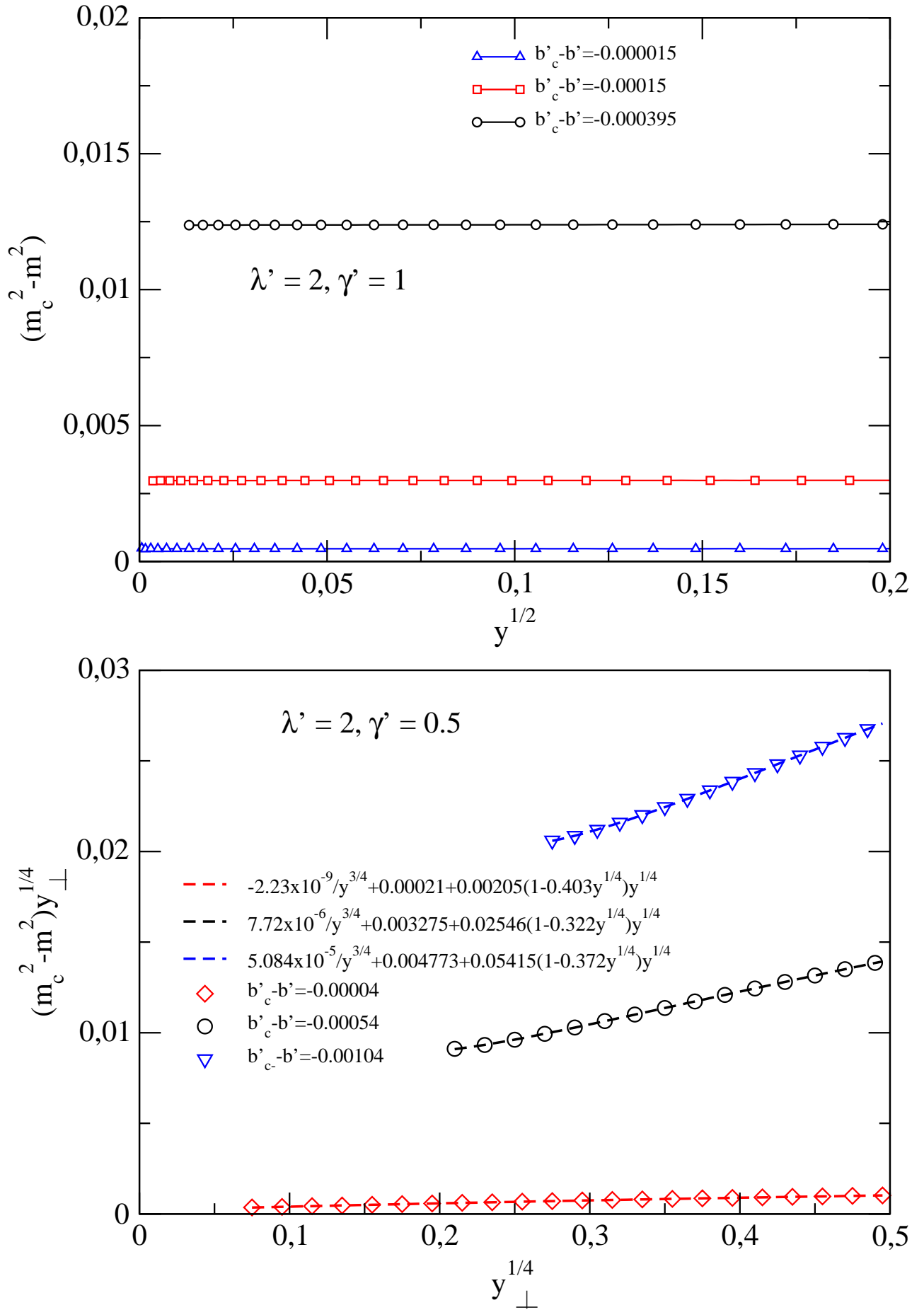}
\caption{Fit of the numerical solution of the HRT equation for D-dimensional spins, in the spherical model limit $\gamma'=1$ and for $\gamma'=0.5$. For $\gamma'=1$ the $m_c^2-m^2$ is represented as function of   $y_\perp^{1/2}$,   while for $\gamma'=0.5$  $(m^2-m_c^2)/y_\perp^{1/4}$ is represented as function of $y_\perp^{1/4}$. In both instances  the sharp cutoff $\lambda'=2$ is considered. Symbols correspond to the numerical solution of the differential equation and the curves denote the results of the  fits.}
\label{fig2}
\end{figure}

In Fig.~\ref{fig2} we plot $(m_c^2-m^2)$ vs $y^{1/2}$ for $\gamma'=1$ (spherical model) and $(m_c^2-m^2)y^{1/4}$ vs $y^{1/4}$  for $\gamma'=0.5$. We consider  various values of $b'$ approaching the critical $b_c'$. In the case of $\gamma'=1$ we find horizontal lines, in accordance with the analytic result (\ref{29}). Thus, in expression (\ref{fit1}) only the coefficient $d$ takes non-zero values while $f$, $c$ and $g$ all vanish. But for $\gamma'=0.5$ all coefficients can be non-zero and the non-linear fits are practically perfect ($r\approx 0.999999$). Additionally,  we have checked that within the expected numerical accuracy, the $t$-dependence of the coefficients also follows  Eq.~(\ref{fit1}). On the other hand, the fact that the coefficients are small can be attributed to the vicinity of the spherical model. Additionally, when compared to the fits of Ref.~\cite{hoye11} (see Fig. 3 therein), one must bear in mind that here we focus on $y_\perp$, while for $D=1$ in Ref.~\cite{hoye11} the $y_\|$ was considered. From our numerical evaluation and analysis in the present work, we have realized that $y_\perp$ can be more easily fitted to expression (\ref{fit1}) for any D (or $\gamma'$), where in the limit $D\rightarrow\infty$ only the coefficient $d$ is retained. With $y_\|$, however, we found that $m_c^2-m^2 = dt -ht^2/\sqrt{y_\|}+\ldots$, where the last term is not included and thus does not fit into   the terms exhibited   in expression (\ref{fit1}). This implies a more complex crossover behavior for $y_\|$ when $\gamma'\rightarrow 1$.

\begin{figure}[H]
\includegraphics[width=17cm,clip]{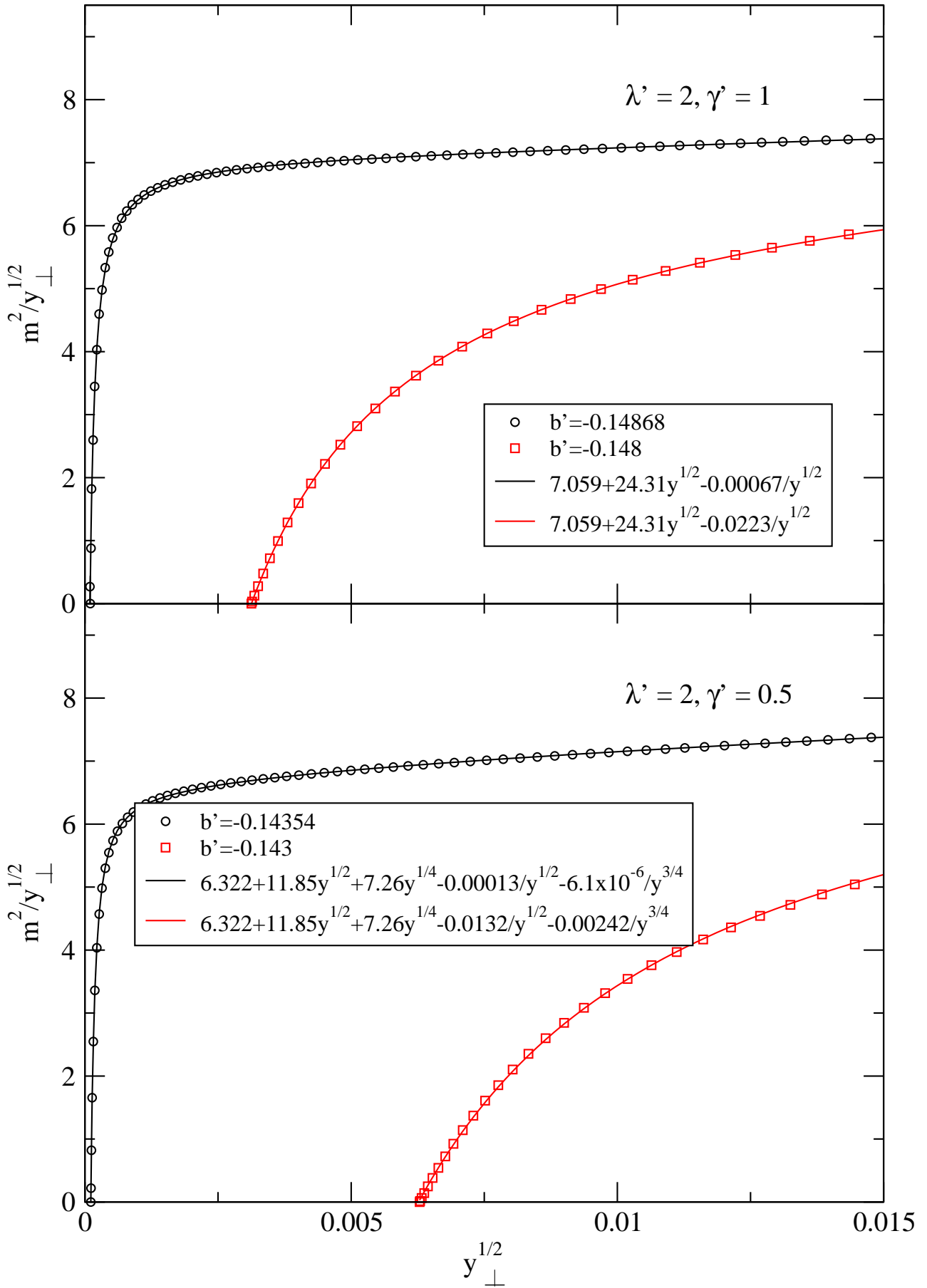}
\caption{Fit of the numerical solution of the HRT equation for D-dimensional spins, in the spherical model limit and for $\gamma'=0.5$, for $m^2/y_\perp^{1/2}$ vs $y_\perp^{1/2}$ for the sharp cutoff $\lambda'=2$. Symbols correspond to the numerical solution of the differential equation, and the curves denote the results of the non-linear fit. The temperature dependence of the fit  is picked up by the coefficients of the last term for the $\gamma'=1$ case and by the last two terms for $\gamma'=0.5$.}
\label{fig3}
\end{figure}

Following this analysis, we now concentrate in those terms that do not depend on temperature and thus represent the critical isotherm. The fit can be found in Figure \ref{fig3} for the spherical model limit (upper graph) and the D-spin case $\gamma'=0.5$ (lower graph). We observe that the spherical model numerical results again comply with the functional form of exact solution (for small $y_\perp$)
\begin{equation}
m^2 = a y_\perp^{1/2} + by_\perp +ct 
\end{equation}
for temperatures quite close to the critical and even somewhat removed from that. The value of the coefficient $a=7.050$ slightly deviates from the exact value  for $\lambda'=2$, $\sqrt{5\pi} \approx 7.025\ldots$, due to the fact that the fit is performed somewhat away from the critical temperature. For the D-spin case with $\gamma'=0.5$, we have as in Ref.~\cite{hoye11} and from Eqs. (\ref{101}) and (\ref{102}), a fit with leading (first and fourth terms respectively) and subleading contributions, which can be expressed as
\begin{equation}
m^2 = a y_\perp^{1/2} + by_\perp +a_1y_\perp^{3/4}  + ct/y_\perp^{1/4} + dt
\label{eosDs}
\end{equation}
where again when compared with the spherical model $\gamma'=1$ case, as found in Ref.~\cite{hoye11}, an additional term $y^{3/4}$ must be added (see the discussion below Eq.(33) in that reference). This can also be expected from the discussion and conclusions   drawn from expressions (\ref{40}) above and (\ref{60}) below. The $t$-dependence   is now collected in the last two terms. One sees in the lower graph of Figure \ref{fig3} that the fit is very accurate, and it can be shown that the dependence on $b'-b'_c$ of the $ct$ and $dt$ coefficients is linear with good accuracy.  It should be noted that the most accurate representation of the $t$-dependence and scaling behavior in the critical region can be extracted from Eq.~(\ref{fit1}). To do so the precise numerical values for the critical isotherm represented by $m_c^2$ (or its expression (\ref{eosDs}) for $t\rightarrow 0$) has been subtracted. In this way expression (\ref{fit1}) is able to determine or fit two additional $t$-dependent terms compared to expression (\ref{eosDs}), i.e. both $f\neq 0$ (a leading scaling term) and $g\neq 0$ can be determined.

\begin{figure}[H]
\includegraphics[width=17cm,clip]{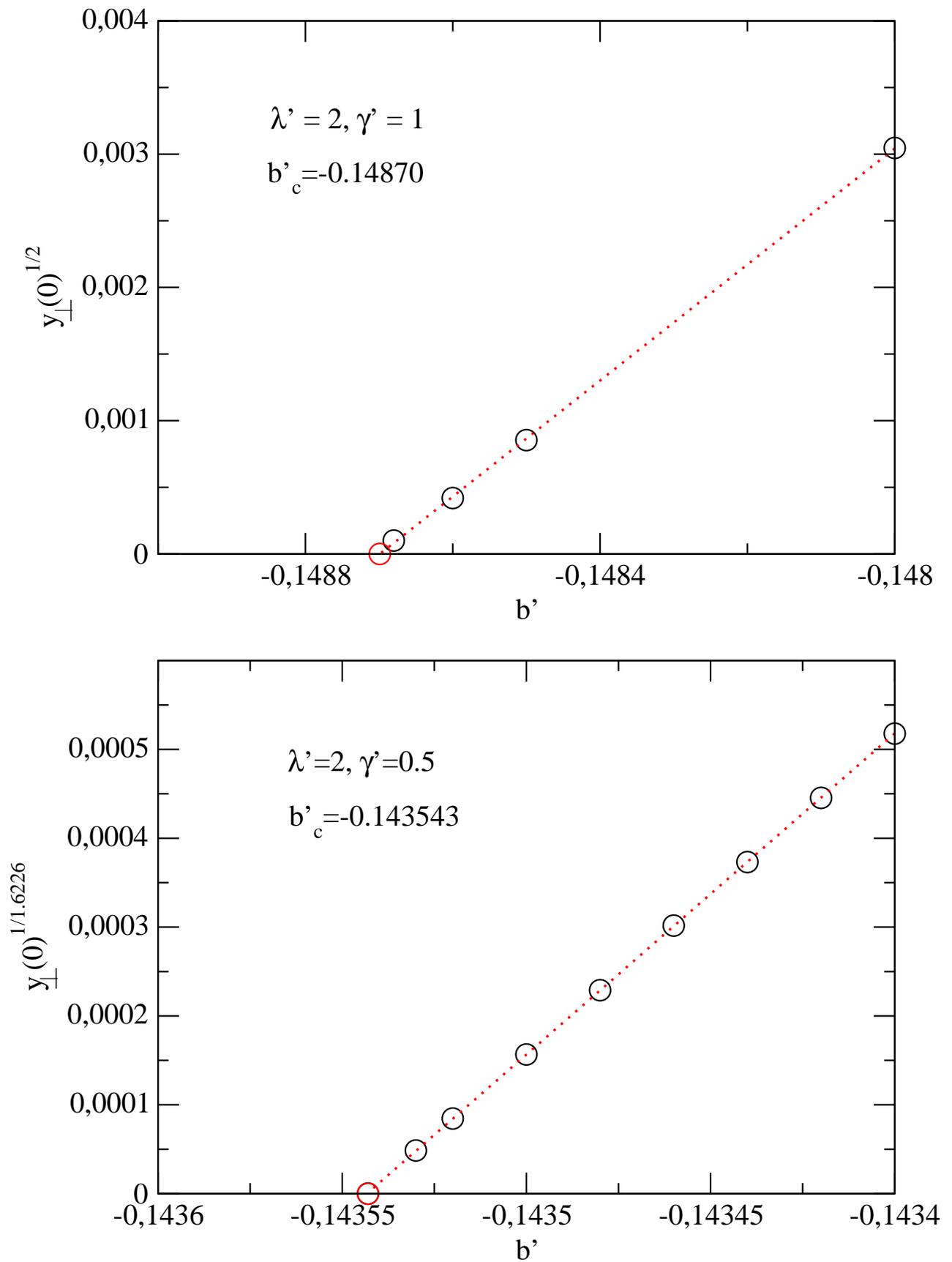}
\caption{Graphical determination of the the critical temperature (critical $b_c'$) for the spherical model limit ($\gamma'=0$) and for the D-dimensional spin case with $\gamma'=0.5$. In the spherical model case the critical index $\gamma=2$, as reflected by the linear dependence of $y_\perp^{1/2}$. For $\gamma'=0.5$ an effective exponent $\gamma = 1.6226$ is found by such a plot.}
\label{fig4}
\end{figure}

\begin{figure}[H]
\includegraphics[width=17cm,clip]{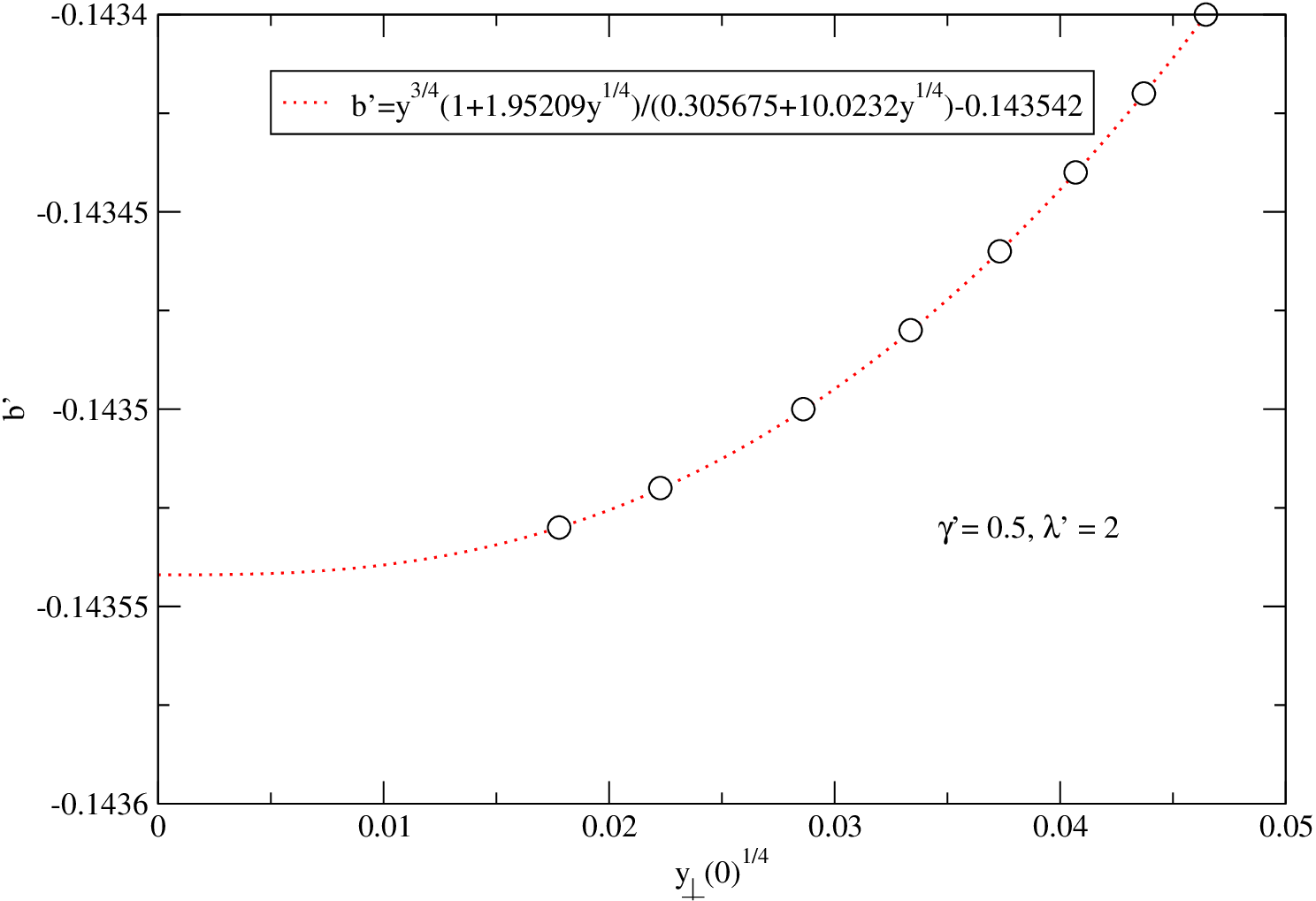}
\caption{Graphical determination of the the critical temperature (critical $b_c'$)  for the D-dimensional spin case with $\gamma'=0.5$ using the crossover expression (\ref{by}) that follows from Eq.~(\ref{fit1}). As indicated in the text, the independent term corresponds to the critical $b'$, in this case $b'_c=-0.143542$. The slight difference with respect to the $b'_c$ obtained in Figure \ref{fig4} reflects the fitting to different functional forms. Symbols correspond to the numerical solution of the differential equation and the curves denote the results of the non-linear fit.}
\label{fig5}
\end{figure}

Now we can test the ability to recover the critical exponent $\gamma$ directly from the numerical calculation. In Ref.~\cite{hoye11} we saw that the numerical solution for $\gamma'=0$ (i.e. D$=1$ spin case) yielded $\gamma=4/3$ in agreement with the HRT analysis both for the sharp and smooth cut-off. Here we see that in the spherical model limit $\gamma'=1$ (upper graph in Fig.~\ref{fig4}) the correct exponent $\gamma=2$ is reproduced. Once the $\gamma'$ departs from 1 we enter the HRT D-spin regime that ends up at $\gamma' = 0$ (1-D case \cite{hoye11}) with $\gamma=4/3$. In the lower graph of Fig.~\ref{fig4} we see that one apparently can obtain an effective $\gamma \approx 1.6226$. However, according to the analysis and the results that follow from Sections \ref{sec3} and \ref{sec6} below, it should   instead   follow   Eq.~(\ref{eosDs}) for $m^2=0$. Thus, with $t=b'_c-b'$ one should then have a function with a crossover regime   between leading and subleading contributions   in which the $t$-dependence of $y_\perp(0)$ when approaching the critical point  can be fitted to   (when neglecting the term $by_\perp$ of Eq.~(\ref{eosDs}), $y=y_\perp$)
\begin{equation}
b'=y^{3/4} \frac{1+c y^{1/4}}{a+b y^{1/4}} + b_c'
\label{by}
\end{equation}
where $a$, $b$ and $c$ are coefficients that follow   from the ones of   Eq.~(\ref{eosDs}).   In the limit $\gamma'=1$ one gets the spherical model with $a=c=0$. In Fig.~\ref{fig5} one observes that the numerical solution for $\gamma'=0.5$ agrees extremely well with the functional dependence expressed in Eq.~(\ref{by}). This explains why, even if  according to our analysis the true critical indices of the HRT for the D-spin model do not depend on D ($\gamma' \neq 1$), one can find apparent effective exponents that depart continuously from the spherical model value.   Thus in Table 2 of Ref.~\cite{Parola2012} it is found that ($\gamma=2\nu$ for $\eta=0$) $\gamma=$1.378, 1.536, and 1.652 for $D=n=$1, 2, and 3 respectively.
Our effective exponent from Fig.~\ref{fig4} for $\lambda'=2$ and $\gamma'=0.5$ ($D=1+\sqrt{5}=3.231$) is similarly $\gamma=1.6226$.  It may be noted that the extrapolations in Figs. \ref{fig4} and \ref{fig5} give slightly different values of $b_c'$.

 In this respect, it is striking how data in a limited region can be well fitted with various types of functions that may influence conclusions drawn based upon their analysis. This is connected to the number of free parameters present. Thus   a power law   function can fit perfectly into 3 data points, whereas   expression (\ref{by})   can fit into 4. 

After these considerations on the properties of the numerical solution of the HRT for the D-spin case, we can now analyze in some more depth its critical  exponents.

%66666666666666666666666666666666666666666666666666666666666666666666666666666666666666666666
\section{Critical indices}
\label{sec6}

Supported by the numerical results we can pursue in more detail the analytic arguments for the precise exponents of the leading scaling function and its subleading contributions. Let us consider the scaling forms for $X$ and $y=y_\|,y_\perp$ given by Eq.~(\ref{40}). With these we can write
\begin{eqnarray}
\nonumber
X^{\prime\prime}&=&\frac{m^4}{t}(X_0+m^\lambda X_1+m^2X_2+\cdots)\\
y_\beta&=&\frac{m^4}{t}(Z_0+m^\lambda Z_1+m^2Z_2+\cdots)
\label{60}
\end{eqnarray}
where  $X_i$ and $Z_i$ ($i=0,1,2$) are new scaling functions in the variables $z=m/Q^{1/2}$ and $\tau=t/Q^{\lambda_1}$. For simplicity the same notation as in Eq.~(\ref{40}) is used, although the functions will be different. Inserted in Eq.~(\ref{34}) for $y_\beta $ this obviously   means   the identities $Z_i=-X_i$ which give no constraints. However, from expression (\ref{33}) for $X$ and Eq.~(\ref{43}) one has that  
\begin{equation}
X\sim J\sim\sqrt{y}\quad(\sim m^2\sim Q).
\label{61}
\end{equation}

  Like Eq.(\ref{43}) the $J$ will contain only this leading contribution    $\propto \sqrt{y}$ as long as $L$ is restricted to the scaling form (\ref{22}).
In addition the $X$ contains $\nu$. But according to the discussion below Eq.~(\ref{39}) the $\nu$ can only contain the scaling terms of $y$ by which the $\nu$ will not modify Eq.~(\ref{61}). So when expanding the $X\sim\sqrt{y}$ in the scaling terms of $y$ we will have (with new scaling functions $Z_i$)
\begin{eqnarray}
\nonumber
X&=&m^2(Z_0+m^\lambda Z_1+m^2 Z_2+\cdots)\\
X^{\prime\prime}&=&Z_0+m^\lambda Z_1+m^2 Z_2+\cdots.
\label{62}
\end{eqnarray}
Expressions (\ref{60}) and (\ref{62}) for $X^{\prime\prime}$ can now be compared. From this it must be concluded that $Z_0=0$    (in Eq.~(\ref{62}))   since $t$ does not scale as $m^4$ at a regular critical point. Thus the $X_0$ and $X_1$ terms must coincide with the $Z_1$ and $Z_2$ terms respectively by which
\begin{equation}
\frac{m^4}{t}\sim m^\lambda \quad \mbox{and} \quad \frac{m^{4+\lambda}}{t}\sim m^2.
\label{63}
\end{equation}
From these expressions one finds $t\sim m^{4-\lambda}$ and $m^{2\lambda}\sim m^2$ or
\begin{equation}
\lambda=1,\quad t\sim m^3\sim  Q^{\lambda_1},\quad \mbox{i.e.}\quad \lambda_1=\frac{3}{2}.
\label{64a}
\end{equation}
With this we obtain simple rational numbers for the critical indices. We have
\begin{eqnarray}
\nonumber
y\sim m^4\sim m^{\delta-1}, \quad m\sim t^{1/3}\sim t^\beta\\
y_\beta\sim\frac{m^4}{t}\sim t^\gamma, \quad \frac{X}{t}\sim\frac{m^6}{t^2}\sim t^{-\alpha}
\label{64}
\end{eqnarray}
by which 
\begin{equation}
\delta=5,\quad\beta=\frac{1}{3},\quad\gamma=\frac{4}{3},\quad\alpha=0,\quad\eta=0,\quad\mbox\quad\nu=\frac{2}{3}.
\label{65}
\end{equation}
The $\alpha$ is the specific heat exponent, while $\eta$ and $\nu$ are the indices for the power law decay of the correlation function and its exponential decay respectively. The analysis above should also hold for $D$-dimensional spins with $D>1$. For subcritical temperatures it is also expected that the corresponding critical indices will be the same, at least for D=1, i.e. $\gamma'=\gamma$, $\alpha'=\alpha$, and $\nu'=\nu$. It can be remarked that the expected subcritical indices are based upon arguments made in Sec.~VIII of Ref.~\cite{hoye09}. 
However for $D>1$ this will be modified somewhat since then horizontal isotherms are expected at phase coexistence as is the case for the MSM and GMSM ($D=\infty$). As already mentioned above, this horizontal slope was found by numerical investigation of SCOZA \cite{pini02} for $D=3$. For this case the critical exponent for the curve of coexistence became the one of the MSM, $\beta=1/2$, within numerical accuracy. This contrasts with the SCOZA for $D=1$ with its generalized scaling that gave \cite{borge98,hoye00} $\beta=0.35$.

Expressions (\ref{60}) and (\ref{62}) may be commented a bit further. 
For the MSM, GMSM, and coinciding HRT and SCOZA of Sec.~\ref{sec4} the $m^\lambda$ terms are not present. Then the $X_0$ and $Z_2$ terms   of Eqs.~(\ref{60}) and (\ref{62}) respectively should be compared to obtain
\begin{equation}
\frac{m^4}{t}\sim m^2, \quad m^2\sim t,\quad\mbox{or}\quad \beta=\frac{1}{2}
\label{66}
\end{equation}
It is also possible to relate $X_0$ to a non-zero $Z_0$   (Eq.~(\ref{62})). Then
\begin{equation}
\frac{m^4}{t}\sim 1, \quad m^4\sim t,\quad\mbox{or}\quad \beta=\frac{1}{4},
\label{67}
\end{equation}
by which   $Z_0=\rm{const}$, $X\propto m^2$, and y$\propto m^4 \propto t$. This corresponds to a tricritical point, and it was
the situation found by the investigation of the SCOZA properties in Ref.~\cite{hoye85}   where a solution with full scaling was found.   Change of boundary conditions (\ref{28}) and (\ref{mf}) into $y_\perp=a' m^4+b'$ will give this situation.
\begin{equation}
\label{}
\end{equation}

%8888888888888888888888888888888888888888888888888888888888888888888888888888888888888888888888
\section{Summary}
\label{sec7}

The HRT for spin dimensionality $D$ has been established and its critical region for supercritical temperatures has been analyzed. From its close connection to SCOZA it is   found   that subleading contributions to the leading scaling behavior  are   important for the critical behavior and the resulting critical indices.   Thus in Sec.~\ref{sec3} we relate the HRT to the corresponding SCOZA problem with an extra parameter $\nu$. There it is shown that this parameter $\nu$ must have the same scaling terms as follows from the ones of $y$. In view of the thermodynamic self-consistence of the SCOZA this implies a connection between a leading and two levels of subleading terms to obtain regular scaling.  

Based upon this and numerical evaluations we find that within numerical accuracy the critical indices of the HRT are simple rational numbers independent of   $D$ $( < \infty)$ as given by Eq.~(\ref{65}).  However, from numerical evaluation alone it would be very difficult to notice this independence. This is clearly demonstrated when comparing Figs.~\ref{fig4} and \ref{fig5} for   $\gamma'=0.5$ where plots with effective exponent and scaling expression (with leading and subleading contributions) respectively are used.   According to our analysis Eq.~(\ref{by}) is the correct one for the HRT, and as shown in Fig.~\ref{fig5} it fits numerical data very accurately. Nevertheless the assumption of an effective simple power law easily fits into the data as shown in Fig.~\ref{fig4}. 
Thus from our analysis as supported by numerical data, we find no crossover behavior with respect to critical indices. As mentioned in  Sec.~\ref{sec1}, this contrasts and is not in agreement with the conclusions drawn from HRT evaluations of Ref.~\cite{Parola2012} and thus not   with $1/n$ or $1/D$ expansions \cite{fisher72}. Instead there is a crossover like the one in Eq.~(\ref{by}) for the susceptibility where the coefficients $a$, $b$, and $c$   in that equation   obviously will vary with $D$ (where $a=c=0$ for $D=\infty$). According to our analysis, a change of $\gamma'$ will thus not change the qualitative behavior of our results, so numerical results for other values of $\gamma'$ (besides $\gamma'=1$ or $D=\infty$) are not included in this work.
  Also it can be noted that expression (\ref{eosDs}) together with (\ref{fit1}) with coefficients determined via numerical results, give an explicit and accurate representation of the equation of state via the susceptibility in any direction away from the critical point. This is seen from Figs.~\ref{fig2} and \ref{fig3}. Thus both leading and subleading scaling contributions are crucial to describe critical properties accurately.

%AAAAAAAAAAAAAAAAAAAAAAAAAAAAAAAAAAAAAAAAAAAAAAAAAAAAAAAAAAAAAAAAAAAAAAAAAAAAAAAAAAAAAAAAAAAAAAa

\section*{Acknowledgments}
E.L. gratefully acknowledges the support from the Direcci\'on
General de Investigaci\'on Cient\'{\i}fica  y T\'ecnica under Grant
No. FIS2010-15502 and from the Direcci\'on General de
Universidades e Investigaci\'on de la Comunidad de Madrid under
Grant No. S2009/ESP/1691 and Program MODELICO-CM.

%BBBBBBBBBBBBBBBBBBBBBBBBBBBBBBBBBBBBBBBBBBBBBBBBBBBBBBBBBBBBBBBBBBBBBBBBBBBBBBBBBBBB

\newpage

\end{document}